\documentclass[iop]{emulateapj}

\usepackage{amsmath, amsthm, amssymb, amsfonts}
\usepackage{hyperref}
\hypersetup{
	colorlinks=true,
	linkcolor=black,
	filecolor=black,
	urlcolor=black,
	citecolor = blue,
}

\def\approxprop{%
  \def\p{%
    \setbox0=\vbox{\hbox{$\propto$}}%
    \ht0=0.6ex \box0 }%
  \def\s{%
    \vbox{\hbox{$\sim$}}%
  }%
  \mathrel{\raisebox{0.7ex}{%
      \mbox{$\underset{\s}{\p}$}%
    }}%
}

\slugcomment{DRAFT VERSION OCTOBER 29, 2019}

\shorttitle{Angular momentum and Morphological Sequence of Massive Galaxies through {\sc Dark Sage}}
\shortauthors{Porras-Valverde et al.}

\begin{document}


\title{Angular momentum and Morphological Sequence of Massive Galaxies through {\sc Dark Sage}}



\author{Antonio J. Porras-Valverde\altaffilmark{1}, Kelly Holley-Bockelmann\altaffilmark{1}, Andreas A. Berlind\altaffilmark{1}, and Adam R. H. Stevens \altaffilmark{2}}
\affil{Vanderbilt University, Department of Physics and Astronomy, Nashville,
TN 37240, USA}
\affil{International Centre for Radio Astronomy Research, The University of Western Australia, Crawley, WA 6009, Australia}







\begin{abstract}

We study the present-day connection between galaxy morphology and angular momentum using the {\sc Dark Sage} semi-analytic model of galaxy formation. For galaxies between $ 10^{11}-10^{12} \mathrm{M}_{\odot}$ in stellar mass, the model successfully predicts the observed trend whereby galaxies with more prominent disks exhibit higher {\em stellar} disk specific angular momentum ($j_{\rm stellar, disk}$) at fixed stellar mass. However, when we include the gas in the disk, bulge-dominated galaxies have the highest {\em total} disk specific angular momentum ($j_{\rm total, disk}$). We attribute this to a large contribution from an extended disk of cold gas in typical bulge-dominated galaxies. We find the relationship between $j_{\rm dark matter}$ and morphology to be quite complex. Surprisingly, in this stellar mass range, not only do bulge-dominated galaxies tend to live in halos with higher $j_{\rm dark matter}$ than disk-dominated galaxies, but intermediate galaxies (those with roughly equal fractions of bulge and disk mass) have the lowest $j_{\rm dark matter}$ of all. Yet, when controlling for halo mass, rather than stellar mass, the relationship between $j_{\rm dark matter}$ and morphology vanishes. Based on these results, halo mass rather than angular momentum is the main driver of the predicted morphology sequence at high masses. In fact, in our stellar mass range, disk-dominated galaxies live in dark matter halos that are roughly 1/10th the mass of their bulge-dominated counterparts. 

\end{abstract}





\section{Introduction}\label{introduction}
For decades, the use of the standard model of cosmology, $\Lambda$CDM, which assumes a flat universe dominated by cold dark matter with a cosmological constant, has greatly contributed to our understanding of galaxy formation \citep{Cen1994X-rayUniverse, Cen1999COSMICEVOLUTION, Cen1999ACCURACYCORRECTIONS, Serna2003, Grande,  Hopkins2009TheDemographics}. It is widely understood that dark matter halos formed from density perturbations in the early Universe, which subsequently acquired angular momentum through tidal torques from neighboring perturbations \citep[see][]{Fields1967, Peebles1969, Harrison}. In a simplistic galaxy formation framework, angular momentum is assumed to be conserved as baryons dissipate energy and quickly collapse into a rotationally-supported disk, but the details of how galaxies form, grow, and acquire angular momentum are far from complete.  Observationally, there is a clear link between the specific angular momentum of stars, $j_{\rm \mathrm{stellar}}$, and the galaxy's stellar mass, where $j_{\rm \mathrm{stellar}}\approxprop \mathrm{M}_{*}^{2/3}$. This relation also carries a well-known dependence on galaxy morphology, where spiral galaxies have higher $j_{\rm \mathrm{stellar}}$ than elliptical galaxies of the same stellar mass \citep{Fall1983, Romanowsky2012, Fall2013, Obreschkow2014, Fall2018ANGULARBULGES}. The relationship between $j_{\rm \mathrm{stellar}}$, stellar mass, and morphology is now often refereed to as the Fall and Romanowsky relation.

There are two techniques for modeling the angular momentum--galaxy connection in a cosmological context: semi-analytic models (SAMs, which use the merger trees of halos from an N-body cosmological simulation to form galaxies within as a post-processing step) and hydrodynamic simulations (which evolve baryons and dark matter together, but at much higher computational expense). A key assumption of most semi-analytic models is that the baryonic specific angular momentum is equal to that of the dark matter through tidal torques \citep{White1991GALAXYCLUSTERING, Cole1994AFormation, Bower2006, Somerville2008, Ricciardelli2010, Benson2012, Croton2016}. As a result, SAMs tend to produce elliptical galaxies in halos with low angular momentum \citep{Avila-Reese1999OnContext, Kauffmann1999Clustering0, Hatton2003GALLICSFormation, Tonini2016TheProcesses}. However, recent cosmological hydrodynamic simulations reveal that the specific angular momentum of baryons can be 3-5 times greater than that of its host dark matter halo \citep{Sales2010, Kimm2011, Pichon2011RiggingDiscs, Stevens2017HowEAGLE}. As one example, this has been attributed to gas-rich mergers and cold flow streams that produce a cool gas disk with high angular momentum that orbits the galaxy halo prior to falling into the galaxy disk at low redshifts \citep{Stewart2011}.

Naturally, the semi-analytic and hydrodynamic techniques must be consistent with observations. Several semi-analytic models have successfully recreated the Fall and Romanowsky relation \citep{Zoldan2018, Zoldan2019}, while others overestimate the total stellar specific angular momentum of disk-dominated galaxies \citep{Mitchell2018}. In both of these semi-analytic models, $j_{\rm \mathrm{baryonic}}$ is coupled to $j_{\rm \mathrm{dark matter}}$ until the halo collapses. As shown by \citet{Stevens2016}, the SAGE model \citep{Croton2016} produces disk-dominated galaxies that overlap observational data from \citet{Fall2013} and \citet{Obreschkow2014}. Although some semi-analytic models agree with observations, these assume $j_{\rm \mathrm{stellar}}$ to be equal to $j_{\rm \mathrm{dark matter}}$. In general, $j_{\rm \mathrm{dark matter}}$ is given by the underlying N-body simulation, whereas $j_{\rm \mathrm{stellar}}$ comes from the prescribed physics from the semi-analytic model, which is indirectly typically constrained by the stellar mass function and often additional observational relations. One model that does not make this simple assumption is {\sc Dark Sage} \citep{Stevens2016}. Instead, {\sc Dark Sage} numerically evolves galaxies' disk structure, meaning, $j_{\rm \mathrm{stellar}}$ is a more physically motivated quantity that accounts for the angular momentum evolution for every galaxy. \citet{Stevens2016} show that {\sc Dark Sage} is able to produce the angular momentum--stellar mass relation for galaxies with a  bulge-to-total stellar mass ratio ($B/T$) less than 0.3. However, they deferred an examination of this relation for galaxies with significant bulge component. Using {\sc Dark Sage}, we want to investigate whether a semi-analytic model that carefully treats angular momentum evolution in disk sizes can reproduce the Fall and Romanowsky relation for massive galaxies within the full bulge-to-total stellar mass range from zero to one. 

An important characteristic of the angular momentum--stellar mass plane is that, at fixed stellar mass, there is a morphological sequence whereby $j_{\rm \mathrm{stellar}}$ increases with decreasing $B/T$ \citep{Posti2018, Sweet2018, Fall2018ANGULARBULGES}. Several studies commonly define galaxy morphology as the ratio of disk-to-bulge stellar mass in a binary form, where a simple cut is used to classify galaxies as disk- or bulge-dominated \citep{Fall2013, Moffett2015, Pedrosa2015, Rosito2018}. There is plenty of evidence from both observations \citep{Matteucci1990METALLICITYBULGE, Kormendy1993KinematicsBulges, Rich1996DidOnce} and simulations \citep{Efstathiou1982, Noguchi1999EARLYDISKS, Croton2006} that bulges can grow through multiple channels, leading to variations in both their spatial structure and kinematic behavior. Several numerical simulations have recreated the angular momentum--stellar mass plane for binary morphology definitions \citep{Zavala2015, Rodriguez-Gomez2016}, but we have yet to see a numerical study that explores this relation using a non-binary morphology sequence. Our galaxy morphology cuts go beyond the separation of galaxies into two morphology bins. In this paper, we show the stellar disk, total disk, and dark matter specific angular momentum for a sequence of galaxy morphology.

Using the observed Fall and Romanowsky relation as our motivation, the goal in this paper is to:
\begin{itemize}
    \item Determine whether the {\sc Dark Sage} semi-analytic model reproduces the observed angular momentum-stellar mass relation for all morphological types
    \item Dissect the angular momentum components such as disk velocity and radius to understand their contribution
    \item Explore the link between galaxy morphology and $j_{\rm \mathrm{stellar, disk}}$, $j_{\rm \mathrm{total, disk}}$, and $j_{\rm \mathrm{dark matter}}$.
\end{itemize}

This paper is organized as follows. In section \ref{Methodology}, we present an overview of the {\sc Dark Sage} semi-analytic model, describe our galaxy sample, and outline our galaxy morphology definition. In sections \ref{Fallrelation} and \ref{halospin}, we examine $j_{\rm \mathrm{stellar, disk}}$, $j_{\rm \mathrm{total, disk}}$, and $j_{\rm \mathrm{dark matter}}$ as a function of stellar mass. Section \ref{discussion} presents our conclusions and discussion about these results.

\section{Semi-analytic Model: {\sc Dark Sage}}\label{Methodology}

{\sc Dark Sage} \citep{Stevens2016} is a model that uses coupled physical and phenomenological analytic expressions to describe the various processes in galaxy formation, leading to predictions of galaxy properties that can be compared to observations. Similar to \citet{Croton2016}, {\sc Dark Sage} halos contain hot gas reservoirs necessary to form and grow galaxies through radiative cooling and condensation of hot gas in halos \citep{White1978}. The hot gas, assumed to be an isothermal sphere \citep{White1991GALAXYCLUSTERING}, cools at a similar rate prescribed in \citet{Croton2016}. The cooling gas collapses within its own gravity to form a galactic disk \citep{Fall1980, Mo1998}. The specific angular momentum of the gas is conserved at first, but subsequent cooling episodes see some angular momentum lost before reaching the disk, as the baryonic and dark matter angular momenta are allowed to decouple. The baryonic angular momentum depends on how unstable it has been over its lifetime, based on its star formation history, gas fraction, feedback processes, and merger history.  

In {\sc Dark Sage}, dark matter halo angular momentum comes directly from the underlying N-body simulation. Over time, the baryonic angular momentum of galaxies decouples from this in the model, as the accretion of gas is typically incoherent; that is, its angular momentum vector is not static (in either its orientation or magnitude). Although the baryonic angular momentum accounts for the galaxy's history, its contribution to its total halo angular momentum is small. In other words, the total halo angular momentum is assumed to be unaffected by the galaxy's history and internal processes.

{\sc Dark Sage} distinguishes itself from other semi-analytic models by how it evolves the radial structure of galaxy disks. {\sc Dark Sage} breaks down disks into 30 equally-spaced bins of specific angular momentum. Every disk has two sets of annuli: one responsible for the stellar and one responsible for the gas angular momentum vectors. Unlike other models of a similar fashion \citep[e.g.][]{Fu2010}, {\sc Dark Sage} does not force these to always be coplanar \citep{Stevens2016}. This becomes important when calculating the two-component Toomre Q stability parameter \citep{Toomre1963}. 

After every cooling episode and merger, {\sc Dark Sage} uses the Q parameter to check stability for each annulus, since internal disk instabilities play an important role in the process of building bulge material \citep{Efstathiou1982, Croton2006, Guo2011, DeLucia2011, Henriques2015} and in the general formation of stars in the model \citep{Stevens2017}. Bulges are built from a combination of secular processes and mergers \citep{Weinzirl2009}. Here, {\sc Dark Sage} allows for matter exchange between annuli from Toomre instabilities, funneling low angular momentum material to the center to grow the bulge component. \citet{Stevens2016} note that by examining galaxies with stable disks, those galaxies naturally follow the observed Fall and Romanowsky relation, showing that the physics considered in {\sc Dark Sage} naturally explain observations. 

To conduct our study, we use the Theoretical Astrophysical Observatory\footnote{https://tao.asvo.org.au/tao/} (TAO) \citep{Bernyk2016TheCatalogues} to construct our data sample. This version of {\sc Dark Sage} is built on merger trees from the Millennium simulation \citep{Springel}. We specifically chose Millennium because {\sc Dark Sage} is well tested on it. The Millennium simulation uses a periodic box with 500 $h^{-1}$Mpc in length, allowing us to sample about $10^5$ galaxies in this paper (see Sec \ref{galaxysample}). It uses cosmological parameters from the Wilkinson Microwave Anisotropy Probe data \citep{Spergel2003} with $\Omega_M = 0.25$, $\Omega_{\Lambda} = 0.75$, $\Omega_b = 0.045$, $\sigma_8 = 0.9$, and $h = 0.73$.\footnote{$\Omega_M$ is the matter density, $\Omega_{b}$ is the dark energy density, $\Omega_{b}$ is the baryon density, and $\sigma_8$ is the amplitude of the linear power spectrum on the scale of 8 $h^{-1}$Mpc} The simulation evolves the dark matter distribution with GADGET-2 \citep{Springel2005} adopting a particle mass of 8.6 x $10^8 h^{-1} \mathrm{M}_{\odot}$. The merger trees are constructed with L-HALOTREE \citep{Springel2005} and the halos and subhalos are found using the SUBFIND \citep{Springel2001}, which also provides the specific angular momentum of each halo. We adopt a minimum halo mass of $10^{11.6} \mathrm{M}_{\odot}$, which corresponds to 400 particles.

It is important to note that {\sc Dark Sage} refers to all ``halo'' quantities as the sum over all matter within the halo. However, because dark matter dominates in mass, we believe that the angular momentum $j$ of the halo is similar to that of the dark matter. For the purposes of this paper, the dark matter, halo, and total $j$ are treated the same.
For a more detailed description of {\sc Dark Sage}, please refer to \citet{Stevens2016}. 

\subsection{Galaxy sample}\label{galaxysample}

\begin{figure}[t]
\centering
\includegraphics[width=\columnwidth, clip]{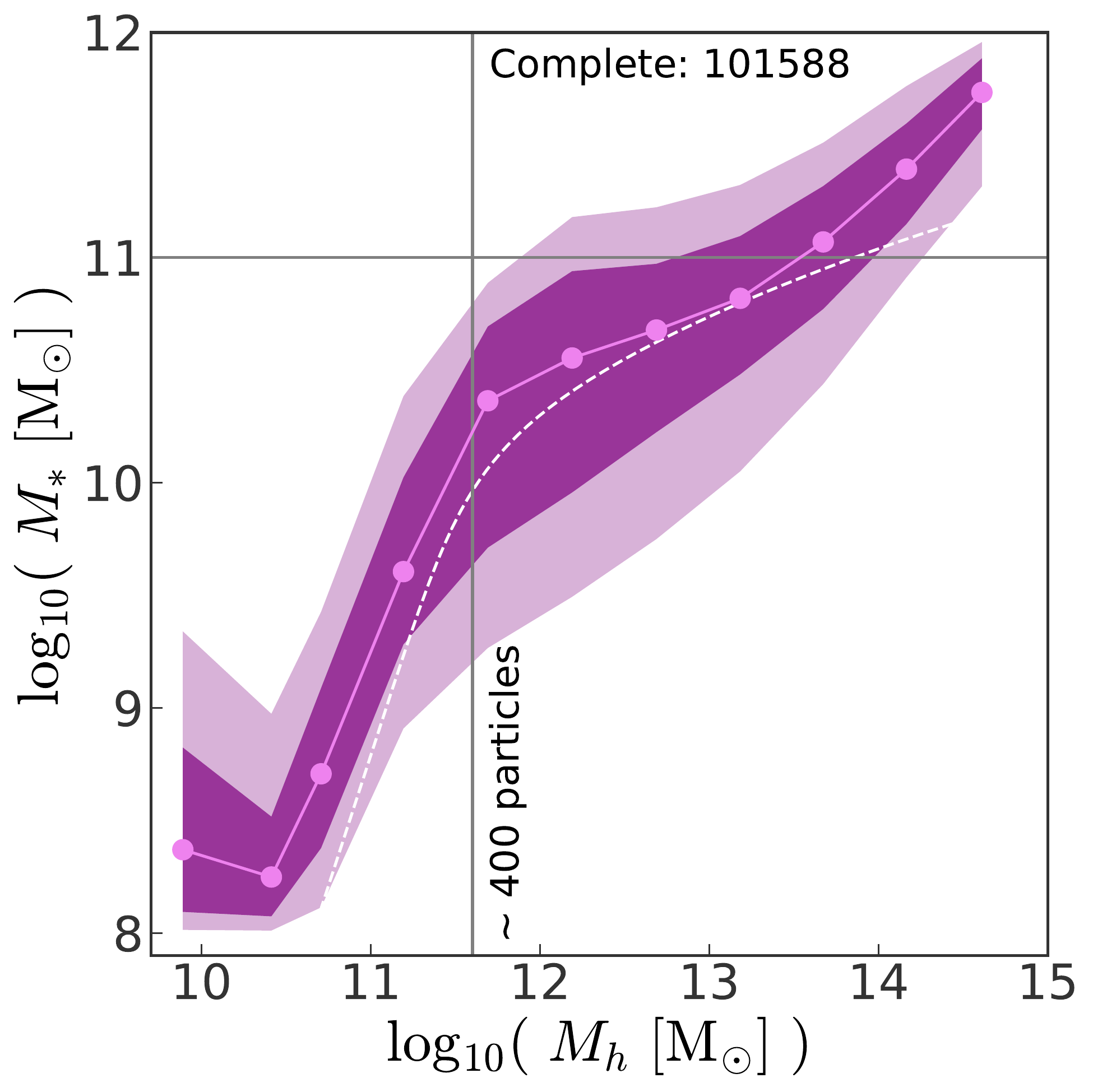}
\caption{ The stellar-to-halo mass relation for central galaxies in {\sc Dark Sage} at z=0. The violet line connecting the dots shows the median binned by halo mass. The dark and light shaded regions enclose 68 and 95 percent of the data, respectively. The white dashed line comes from \citet{Behroozi2010} model. The grey vertical line at $ 10^{11.6} \mathrm{M}_{\odot}$ denotes the halo mass cut, while the grey horizontal line at $ 10^{11} \mathrm{M}_{\odot}$ denotes the stellar mass cut used to conduct our study. Our complete sample contains 101,588 galaxies. }\label{fig:galaxy_sample}
\end{figure}

Our goal is to examine $j_{\rm \mathrm{stellar, disk}}$, $j_{\rm \mathrm{total, disk}}$, and $j_{\rm \mathrm{dark matter}}$ as a function of stellar mass and galaxy morphology. Only central galaxies are part of our study, as the angular momenta of satellites' subhalos in Millennium are not well-resolved. As we mention above, we adopt a halo mass cut of $10^{11.6} \mathrm{M}_{\odot}$, requiring that every halo with a galaxy has at least 400 particles. Since we are interested in a complete sample, we also restrict our data to galaxies with stellar masses between $10^{11}-10^{12} \mathrm{M}_{\odot}$. To illustrate our data selection, Figure \ref{fig:galaxy_sample} shows the stellar-to-halo mass relation for {\sc Dark Sage} galaxies. The violet line represents the median of data binned by halo mass, where all bins include at least 30 galaxies. The dark and light shaded regions enclose the inner 68 and 95 percent of the data, respectively. The white dashed line shows the \citet{Behroozi2010} model, which goes as  $M_h^{2.3}$ at halo masses between $10^{11}-10^{12} \mathrm{M}_{\odot}$ and $M_h^{0.3}$ at halo masses greater than $10^{14} \mathrm{M}_{\odot}$ at redshift 0. The grey vertical line shows our halo mass cut, and the grey horizontal line at $ 10^{11} \mathrm{M}_{\odot}$ denotes the stellar mass cut used to conduct our study. The plot shows that all the galaxies in our sample live in well-resolved halos. Our galaxy sample contains 101,588 galaxies. 

\subsection{Defining galaxy morphology} \label{galaxymorphology}

It is common in observational studies to use luminosity or surface brightness profiles to define galaxy morphology \citep{RobertsMortonS.1994, Blanton, Driver2006, Moffett2015}. \citet{Kauffmann2003} found that massive elliptical galaxies experience high surface mass densities, which usually carry old stellar populations. On the other hand, low-mass galaxies have low surface mass densities, which typically translate to the existence of disks, which commonly contain young stellar populations.

\begin{figure}[t]
\centering
\includegraphics[width=\columnwidth, clip]{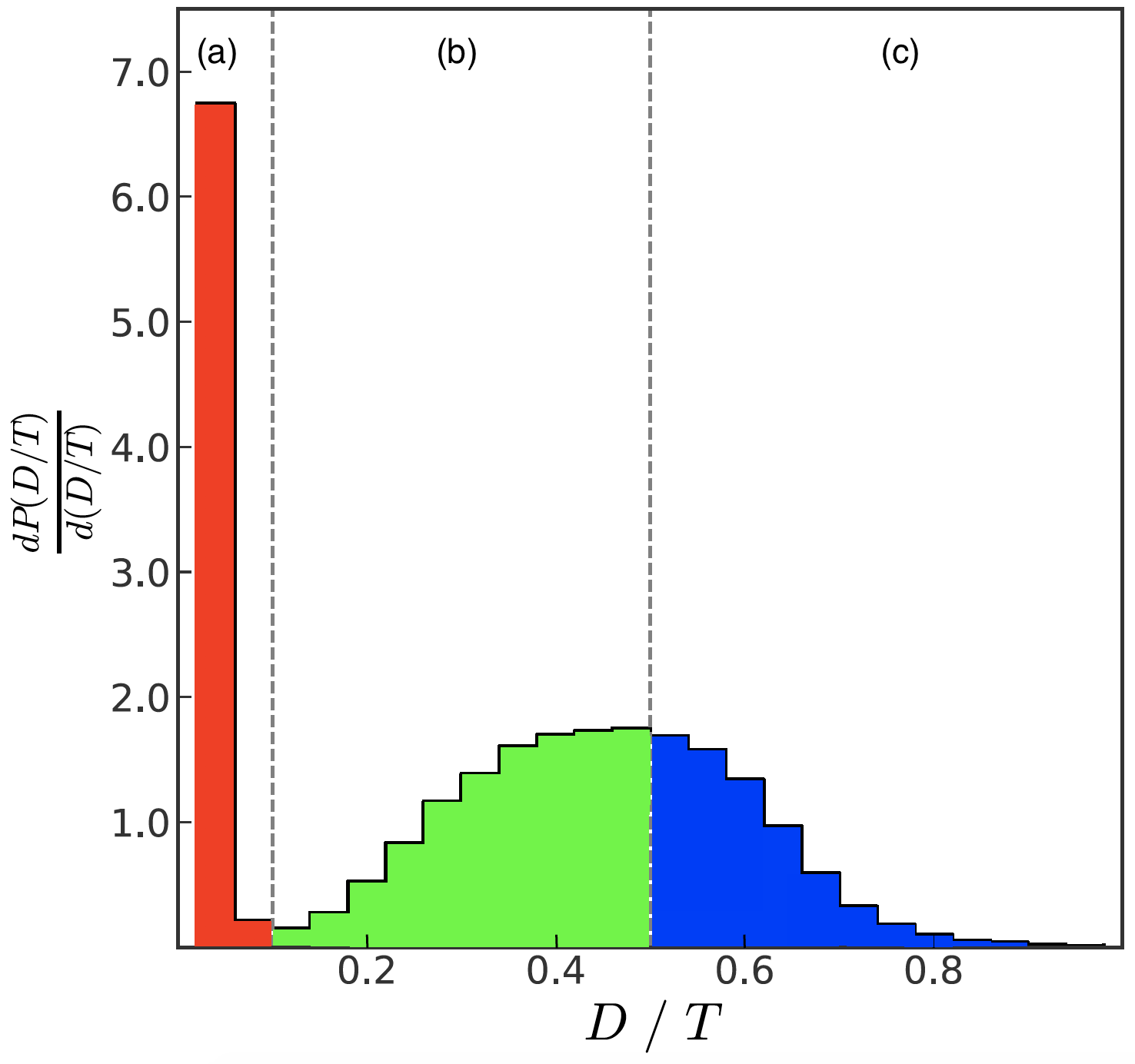}
\caption{Distribution of disk-to-total stellar mass ratio within our galaxy sample at z=0. The vertical black dashed lines denote our morphology cuts. Galaxies with a $D/T$ less than 0.1 are bulge-dominated, containing 28,539 galaxies (Region a; red), galaxies between 0.1 and 0.5 are the intermediate population, containing 48,397 galaxies (Region b; green), and galaxies with a $D/T$ greater than 0.5 are disk-dominated, containing 24,652 galaxies (Region c; blue).}
\label{fig:gal_morph_hist}
\end{figure}

Several theoretical studies define galaxy morphology as the ratio of disk-to-bulge stellar mass in order to separate disk-dominated from bulge-dominated galaxies \citep{Fall2013, Pedrosa2015, Rosito2018}. In our study, galaxy morphology is based on the ratio of disk-to-total stellar mass {$D/T$}, where the total stellar mass is defined as the sum of the disk and bulge stellar mass components. {\sc Dark Sage} treats bulge formation through multiple channels; therefore, the bulge component is divided into merger-driven, instability-driven, and pseudo- bulge components.  The sum of these make up the total bulge mass, which contributes to the $D/T$. Figure \ref{fig:gal_morph_hist} shows the normalized distribution of this ratio for our galaxy sample.
We divide our sample into three morphological types, based on the $D/T$ ratio. Galaxies with $D/T > $ 0.5 are defined as disk-dominated and contain 24652 galaxies (Region c of Figure \ref{fig:gal_morph_hist}).  Galaxies with a $D/T$ between 0.1 and 0.5 are an intermediate population, containing 48397 galaxies (Region b) and galaxies with a $D/T$ less than 0.1 are bulge-dominated, containing 28539 galaxies (Region a). This stellar morphology distribution looks quite similar to that measured by \citet{Bluck2014BulgeSurvey} from the SDSS, which also shows a large narrow peak for massive galaxies with $D/T$ less than 0.1.

\subsection{Calculating angular momentum} \label{AMgalaxies}

{\sc Dark Sage} calculates the total angular momentum inside a disk annulus as: 
\begin{equation}
J_{\mathrm{annulus}} = m_{\mathrm{annulus}} * 0.5 * (j_{\mathrm{inner}} + j_{\mathrm{outer}}),\label{eq:AM_one_annullus}
\end{equation}
where $m_{\mathrm{annulus}}$ is the mass inside the annulus and $j_{\mathrm{inner}}$ and $j_{\mathrm{outer}}$ are the specific angular momenta of the inner and outer boundaries of the annulus, respectively. The sum over the total number of annulli (30 in our case) gives the total angular momentum of the disk structure. Then, the specific angular momentum is obtained by dividing by the total mass of all annulli.

For the disk component, we track the specific angular momentum of the stellar and cold gas components, specifically atomic hydrogen (H{\sc i}) and molecular hydrogen (H2) for the latter:
\begin{equation}
\frac{J_{\mathrm{total, disk}}}{M_{\mathrm{total, disk}}} = \frac{ J_{\mathrm{stellar,disk}} + J_{\mathrm{H}\textsc{i}} + J_{\mathrm{H2}}}{ M_{\mathrm{stellar, disk}} + M_{\mathrm{H}\textsc{i}} + M_{\mathrm{H2}} } = j_{\mathrm{total, disk}},\label{eq:Baryonic_AM}
\end{equation}
where $J_{\mathrm{stellar,disk}}$, $J_{\mathrm{H}\textsc{i}}$, and $J_{\mathrm{H2}}$ are the stellar, atomic, and molecular hydrogen angular momenta within the disk. $J_{\mathrm{stellar,disk}}$ does not include any information about the angular momentum of the bulge component. Also, we consider H{\sc i} and H2 to make up the cold gas. Note that we ignore ionized gas, helium, and gaseous metals in the disk.

\section{Connecting the stellar and total disk specific angular momentum}\label{Fallrelation}

\begin{figure}[t]
\centering
\includegraphics[width=\columnwidth, clip]{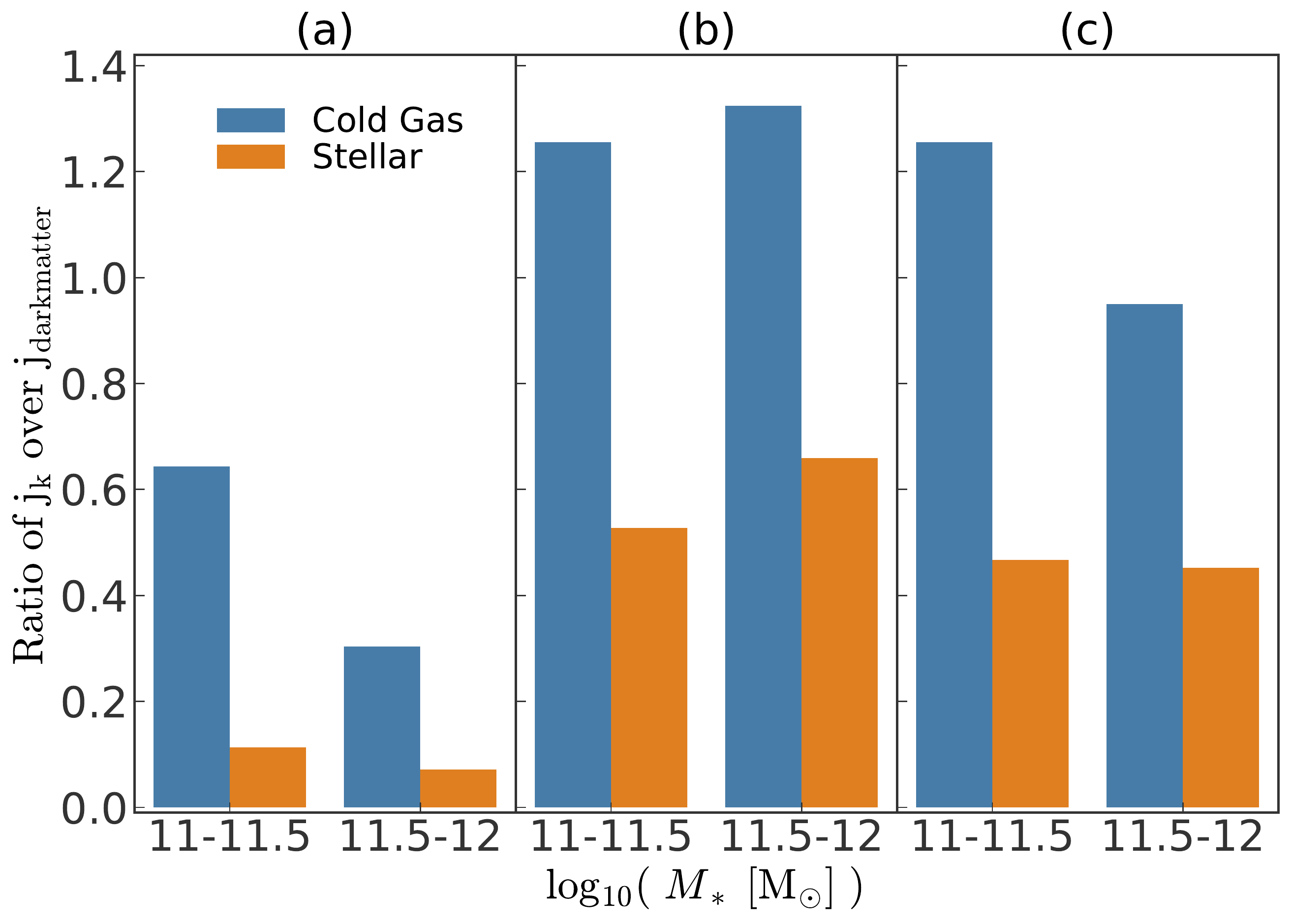}
\caption{From left to right: the ratio of $j_{\mathrm{H\textsc{i}+H2}}$ (blue) and $j_{\mathrm{stellar, disk}}$ (orange) over $j_{\mathrm{dark matter}}$ within the stellar mass  ranges of $10^{11}-10^{11.5}$ and $10^{11.5}-10^{12} \mathrm{M}_{\odot}$ for bulge-dominated (panel a), intermediate (panel b), and disk-dominated galaxies (panel c). For all galaxies, the cold gas component has the most specific angular momentum. Intermediate and disk-dominated galaxies have higher $j_{\mathrm{H\textsc{i}+H2}}$ than $j_{\mathrm{dark matter}}$. Bulge-dominated galaxies have the least $j_{\mathrm{stellar, disk}}$}\label{fig:barplots_morph}
\end{figure}

To dissect the anatomy of our galaxy morphology sequence, in Figure \ref{fig:barplots_morph}, we break down the angular momentum contribution of the stellar and cold gas components. Figure \ref{fig:barplots_morph} illustrates the ratio of $j_{\mathrm{H\textsc{i}+H2}}$ (blue) and $j_{\mathrm{stellar, disk}}$ (orange) over $j_{\mathrm{dark matter}}$ within the stellar mass ranges of $10^{11}-10^{11.5}$ and $10^{11.5}-10^{12} \mathrm{M}_{\odot}$ for bulge-dominated (panel a), intermediate (panel b), and disk-dominated galaxies (panel c).

Figure \ref{fig:barplots_morph} shows that, for bulge-dominated galaxies, the dark matter component has the highest specific angular momentum, followed by cold gas, and the stellar disk coming in last. The dominance of $j_{\mathrm{dark matter}}$ is strongest for galaxies with higher overall stellar mass. In contrast, for intermediate and disk-dominated galaxies, the gas disk generally has the highest specific angular momentum, with dark matter coming next and the stellar disk coming in last. We do not see a strong trend of these results with stellar mass.

\begin{figure}[t]
\centering
\includegraphics[width=1.1\columnwidth, clip]{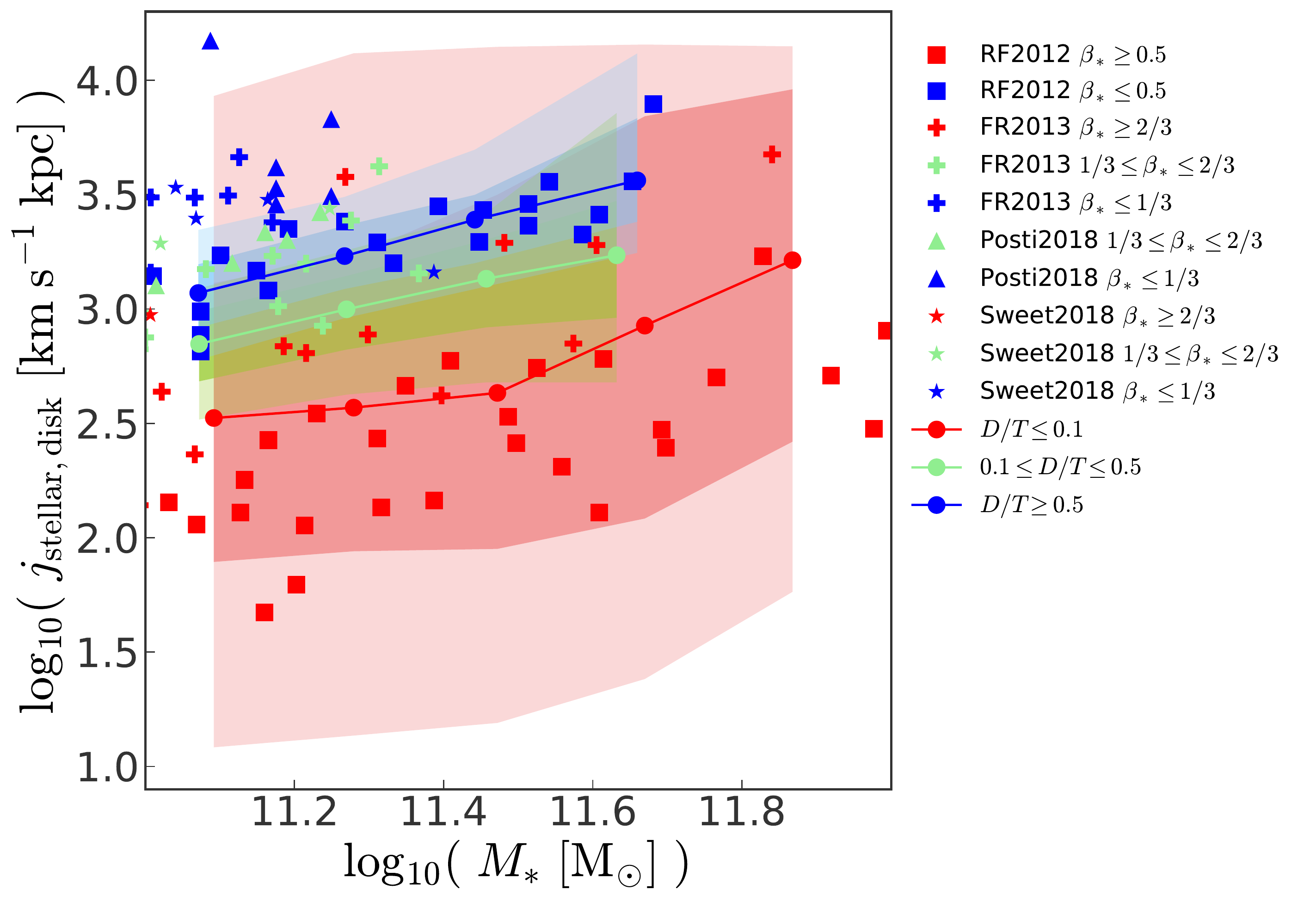}
\caption{Specific angular momentum of the stellar disk as a function of stellar mass for bulge-dominated (red), intermediate (green) and disk-dominated (blue) galaxies. The solid lines show the median within stellar mass bins of width 0.2 dex. The dark and light shaded regions enclose the 68 and 95 percentile of the data for the respective distributions. Squares, crosses, triangles, and stars come from \citet{Fall1983, Romanowsky2012, Fall2013, Posti2018, Sweet2018, Fall2018ANGULARBULGES} observational surveys, as shown in the figure legend, along with the bulge-to-total fractions $\beta$ that correspond to each data point. Dark Sage reproduces the weak trend of $j_{\rm stellar}$ with stellar mass and morphological separation for disk and bulge dominated systems.}
\label{fig:sAM_SMplane}
\end{figure}

We now explore the Fall and Romanowsky relation to test {\sc Dark Sage} against observations. Figure~\ref{fig:sAM_SMplane} shows the specific angular momentum of the stellar disk as a function of stellar mass for bulge-dominated (red), intermediate (green), and disk-dominated (blue) galaxies. The solid lines connecting the dots show the running medians for {\sc Dark Sage} galaxies binned by stellar mass. The dark and light shaded regions enclose 68 and 95 percent of the data from their respective distributions. The squares, crosses, triangles, and stars come from observations~\citep{Fall1983, Romanowsky2012, Fall2013, Posti2018, Sweet2018, Fall2018ANGULARBULGES}.

It should be noted that the observational studies use different bulge and disk decomposition methods, concentrate on different morphological cuts, and span different mass ranges to define their samples. For example, \citet{Romanowsky2012} use a fixed mass-to-light ratio, while \citet{Fall2013} use a variable mass-to-light ratio for the bulge and disk components. This is important to mention given that distinct mass-to-light ratios result in different bulge and disk components.\footnote{Some galaxies are present in several of the observational studies, and are therefore plotted more than once in Figure~\ref{fig:sAM_SMplane}. We opted to allow this, rather than only showing one result per galaxy, in order to remain agnostic as to the different methods used in deriving galaxy properties observationally. In that sense, some of the systematic uncertainties in the observational measurements are accounted for in the the comparison with {\sc Dark Sage}} Galaxies in \citet{Romanowsky2012} and \citet{Fall2013} cover a stellar mass range of $ 10^{8.9}-10^{11.8} \mathrm{M}_{\odot}$ and a range of bulge fractions between 0 and 1. \citet{Romanowsky2012} use bulge fractions ($\beta_*\equiv 1 - D/T$) and define spiral and elliptical galaxies with $0\leq \beta_* \leq 1/2$ and $1/2\leq \beta_* \leq 1.0$, respectively. \citet{Fall2013} and \citet{Sweet2018} also use bulge fraction cuts to define spiral ($0\leq \beta_* \leq 1/3$), intermediate ($1/3\leq \beta_* \leq 2/3$), and elliptical ($2/3\leq \beta_*\leq1$) galaxies. \citet{Sweet2018} use galaxies from the CALIFA survey \citep{Sanchez2012} and fit an exponential disk profile and a S\'ersic bulge to distinguish the disk and bulge components. This sample spans a stellar mass range of $ 10^{9.5}-10^{11.4} \mathrm{M}_{\odot}$ with a range of bulge fractions between 0 and 0.7. Lastly, \citet{Posti2018} use observed surface brightness profiles to define an outer disk component and covers a stellar mass range of $ 10^{7.0}-10^{11.3} \mathrm{M}_{\odot}$ with a narrow range of bulge fractions between 0 and 0.3. Their sample cuts for intermediate and spiral galaxies are $1/3\leq \beta_* \leq 2/3$ and $0\leq \beta_* \leq 1/3$, respectively. We note that our cuts are not the same as the observational cuts. However, the results of {\sc Dark Sage} do not change significantly if we make the same cuts as in observations. We stick with our cuts because they are more natural given the morphology distribution seen in Figure~\ref{fig:gal_morph_hist}.

Figure~\ref{fig:sAM_SMplane} shows that {\sc Dark Sage} successfully reproduces most of the observational trends for our sample. The median relations of disk, intermediate, and bulge-dominated galaxies predicted by {\sc Dark Sage} go through the corresponding observational data points. The full height of the 16-84th inner percentile range scatter of $j_{\mathrm{stellar, disk}}$ for disk-dominated galaxies in {\sc Dark Sage} is 0.3 dex, which is about the same as the observed scatter of disk-dominated galaxies in \citet{Romanowsky2012}. The scatter for bulge-dominated galaxies in {\sc Dark Sage} is 1.2 dex, and increases with stellar mass to 1.6 dex. This is larger than the (full width) scatter observed by \citet{Romanowsky2012}, which is closer to 0.6 dex.

\begin{figure}[t]
\centering
\includegraphics[width=\columnwidth, clip]{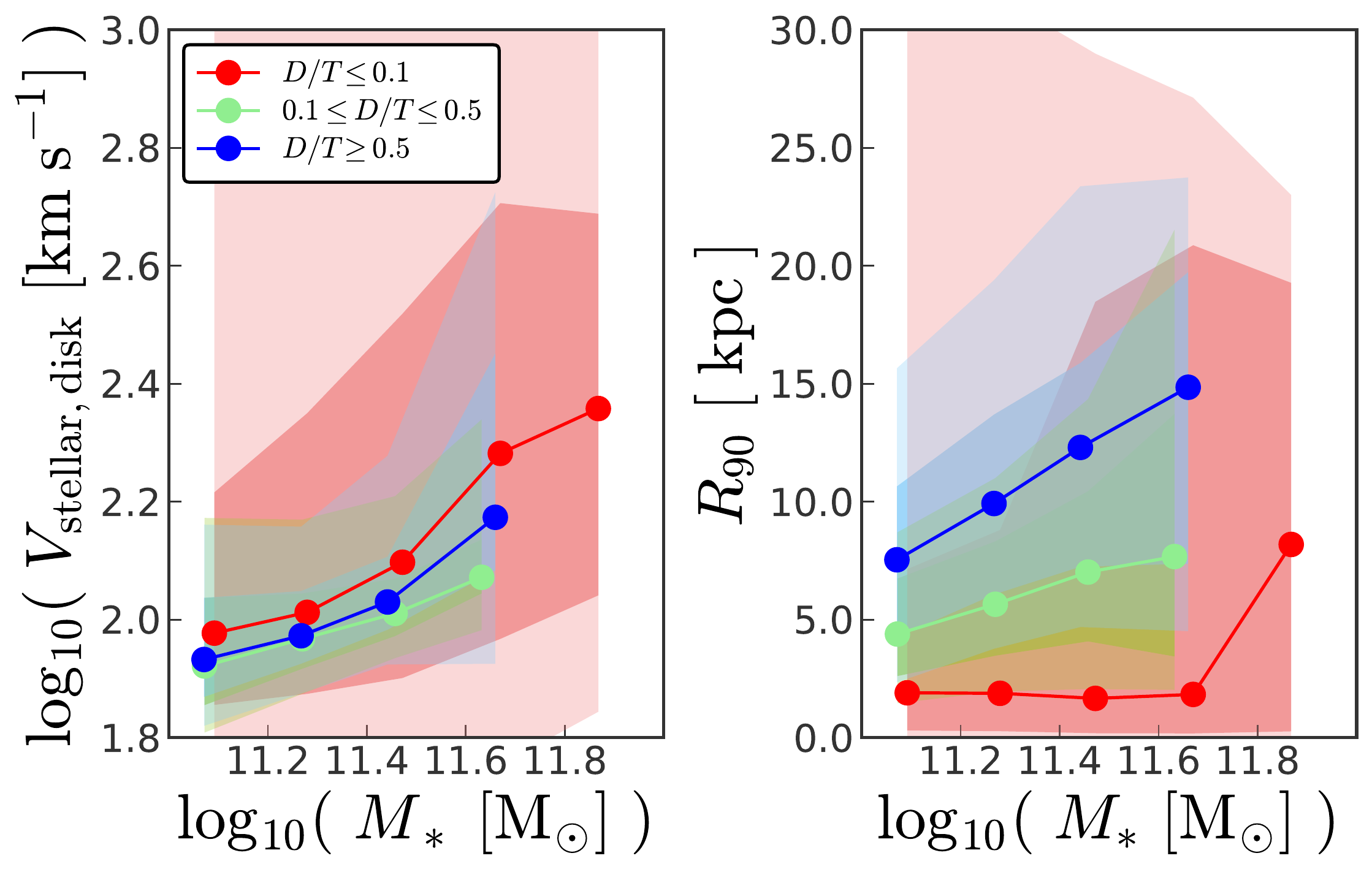}
\caption{Average stellar disk velocity (left panel) and disk radius (right panel) as a function of stellar mass, for bulge-dominated (red), intermediate (green) and disk-dominated (blue) galaxies. Disk radius is the radius ($R_{90}$) of the disk enclosing 90\% of the stellar mass. Bins, lines, and shaded regions are the same as in Fig.~\ref{fig:sAM_SMplane}. All three galaxy populations have similar average disk velocities, but different average disk radii. Generally, disk-dominated galaxies have disk structures that are about two times larger than the disk structures in bulge-dominated galaxies. This difference increases with stellar mass.}\label{fig:Vdisk_SM}

\end{figure}

A different way to investigate the morphological dependence of disk angular momentum and its scatter in {\sc Dark Sage} is to look at disk velocities and radii. Figure \ref{fig:Vdisk_SM} shows the radius, $R_{90}$, of the disk where 90 percent of the stellar mass is enclosed (right panel) and the average velocity, $V_{\mathrm{stellar, disk}}$ = $j_{\mathrm{stellar, disk}}$/ $R_{90}$, of the disk (left panel) as a function of stellar mass. We find that all three galaxy populations have similar median disk velocities and overlaying distributions. The scatter of disk velocity for intermediate and disk-dominated galaxies is about 0.1 dex, slightly increasing with stellar mass, while bulge-dominated galaxies have a much larger scatter of 0.3 dex increasing to 0.6 dex within our mass range.

When looking at the radius of the disk, we find a morphological sequence in which, at a given stellar mass, disk-dominated galaxies have larger disk structures, followed by intermediate galaxies with intermediate sized disks, and bulge-dominated galaxies, which have the smallest disks. In general, disk-dominated galaxies have disk radii that are about two times larger than bulge-dominated galaxies. This difference increases with stellar mass. The scatter (full width) of $R_{90}$ for disk-dominated and intermediate galaxies is about 5 kpc, increasing with stellar mass, while bulge-dominated galaxies have scatter 7 kpc increasing to 20 kpc within our mass range. Note that $R_{90}$ only refers to the size of the stellar disk in the galaxy and it does not make any predictions for the size of the bulge structure. Bulge-dominated galaxies within our sample have a large range of disk sizes. The median of the distribution shows that most massive bulge-dominated galaxies have small $R_{90}$. Nevertheless, there is a large scatter in the distribution, where a significant number of these also have large disk structures. The large scatter in both the velocity of the disk and $R_{90}$ are the reflection of the large variations in the $j_{\mathrm{stellar, disk}}$, which constantly varies based on how major and minor mergers disrupt the structure of the disk. The change in $j_{\mathrm{stellar, disk}}$ is what drives the variation in disk size. 

\begin{figure}[t]
\centering
\includegraphics[width=\columnwidth, clip]{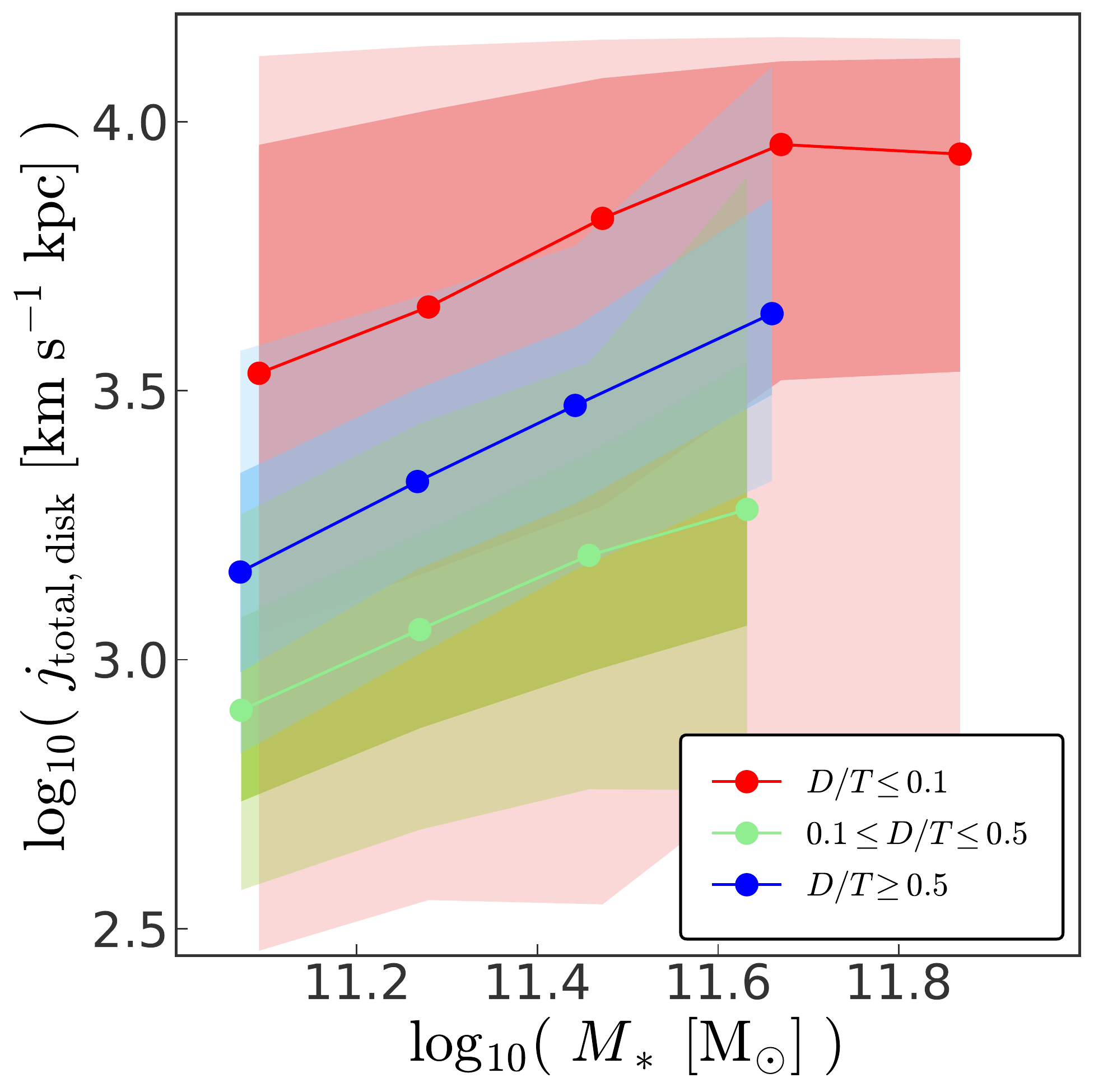}
\caption{Total (stellar + gas) specific angular momentum of the disk as a function of stellar mass for bulge-dominated (red), intermediate (green), and disk-dominated (blue) galaxies. The figure includes running median lines, while the dark and light shaded regions enclose the inner 68 and 95 percent of the sample, respectively (as in Fig.~\ref{fig:sAM_SMplane}). We find that $j_{\rm total disk}$ of disk-dominated and intermediate galaxies tracks $j_{\rm stellar}$. Surprisingly, bulge-dominated galaxies have the highest $j_{\rm total disk}$, implying that $j_{\rm cold gas}$ boosts $j_{\rm total disk.}$ }\label{fig:BaryonicAM_SMplane}

\end{figure}

\citet{Obreschkow2014}, \citet{Butler2017}, and \citet{Wang2018AngularNIHAO} investigated baryonic, rather than stellar, angular momentum, and found a strong empirical correlation between $j_{\mathrm{baryonic}}$, baryonic mass and morphology, whereby galaxies with high bulge-to-total stellar mass ratios have lower $j_{\mathrm{baryonic}}$. We use the cold gas and stellar components within the disk to investigate this with {\sc Dark Sage}. Figure~\ref{fig:BaryonicAM_SMplane} illustrates the $j_{\mathrm{total, disk}}$ (including both the cold gas and stellar component) to stellar mass relation for our galaxy sample. Like in Figure~\ref{fig:sAM_SMplane}, the dots show the median values, and the dark and light shaded regions enclose 68 and 95 percent of the samples, respectively. We find that $j_{\mathrm{total, disk}}$ of disk-dominated and intermediate galaxies track $j_{\mathrm{stellar, disk}}$. Surprisingly however, bulge-dominated galaxies have the highest $j_{\mathrm{total, disk}}$. The scatter of the $j_{\mathrm{total, disk}}$ for disk-dominated and intermediate galaxies is about 0.4 dex and independent of mass, while for bulge-dominated galaxies, the scatter is about 1 dex and decreasing with mass to 0.5 dex.

This contrast to Figure~\ref{fig:sAM_SMplane} arises because bulge-dominated galaxies contain disks with high gas fractions in {\sc Dark Sage}. Gas disks tend to have higher \textit{j} than stellar disks, as disk stars preferentially form from the lowest-\textit{j} gas in the disk (see Fig.~2 of ~\citealt{Stevens2018}).  In bulge-dominated galaxies, these gas disks are typically low mass and insufficiently stable or dense to form stars. As shown in Figure \ref{fig:barplots_morph}, the ratio of $j_{\mathrm{H\textsc{i}+H2}}$ and $j_{\mathrm{stellar, disk}}$ is much higher for bulge-dominated galaxies (around 6) than for intermediate and disk-dominated galaxies (closer to 2).

\section{Dark matter specific angular momentum and Morphology: is there a correlation?} \label{halospin}

There is an expected relationship between galaxies and their dark matter specific angular momenta, as galaxies are believed to initially form through the collapse of gas within the halo (dissipating energy to create a rotationally-supported disk; \citealt{Peebles1969}). Analytic galaxy formation models show that the galactic disks formed from halos with high spin should have large sizes \citep{Mo1998}. This simple idea is at the core of most semi-analytic models and thus provides a strong motivation to better understand how the dark matter halo impacts galaxy morphology. Seemingly contradicting this foundational framework, \citet{Rodriguez-Gomez2016} found little-to-no correlation between the halo spin parameter \citep{Bullock2000} and kinematic morphology (dispersion- vs. rotation- dominated galaxies) within the stellar mass range $10^{11}-10^{12} \mathrm{M}_{\odot}$ for central galaxies at $z=0$ in the high resolution Illustris simulation. At lower stellar mass, however, they do find that rotation-dominated galaxies have a higher halo spin parameter than dispersion-dominated galaxies. \citet{Rodriguez-Gomez2016} use a kinematic definition of morphology, taking a fraction of the kinetic energy invested in ordered rotation \citep{Sales2012}. Using {\sc Dark Sage}, we are able to reproduce similar results for halo spin if we adopt a binary morphology cut. In other words we find that bulge- and disk-dominated galaxies have similar dark halo spin distributions.

\begin{figure}[t]
\centering
\includegraphics[width=\columnwidth, clip]{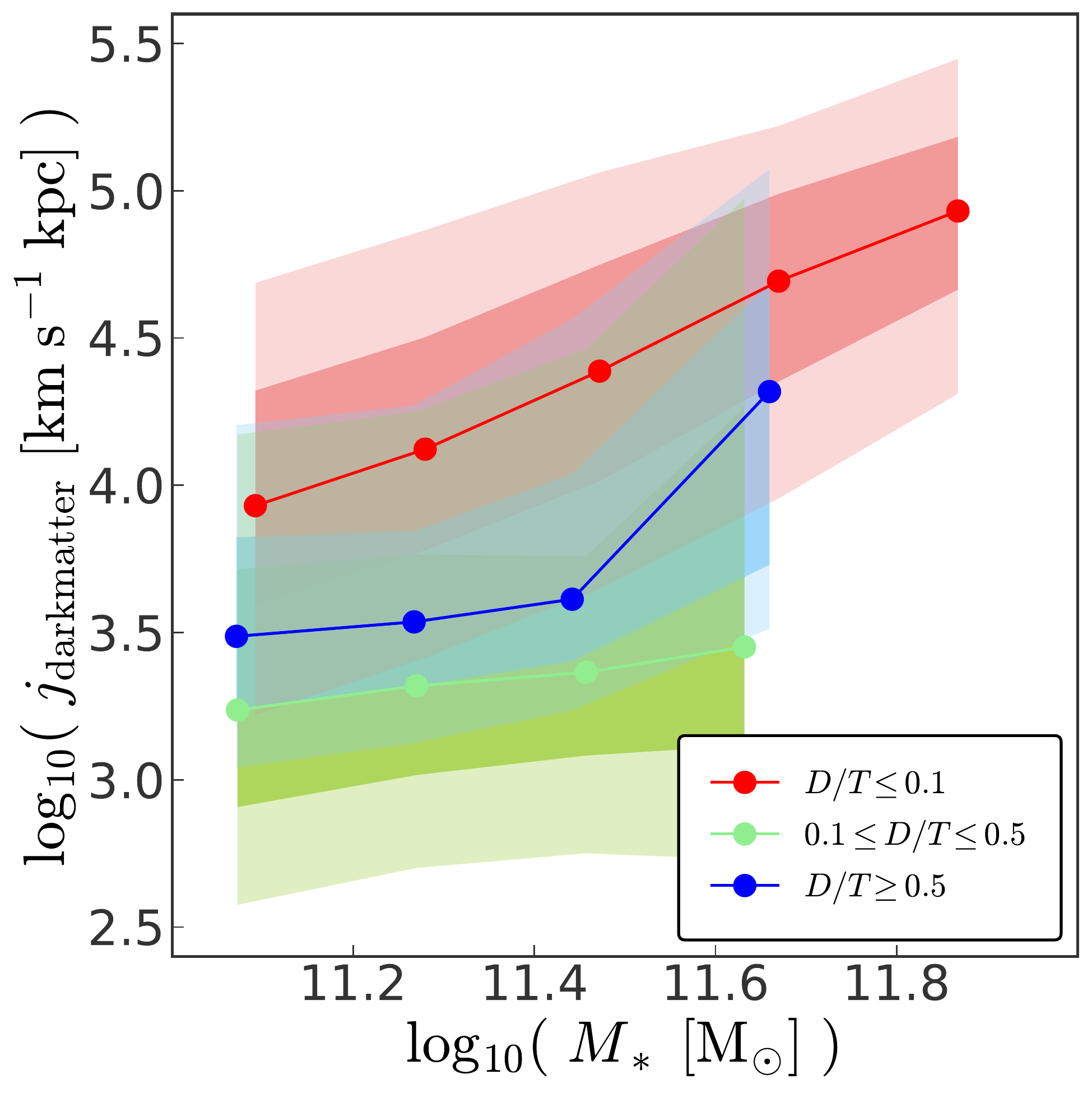}
\caption{Dark matter specific angular momentum as a function of stellar mass for bulge-dominated (red), intermediate (green), and disk-dominated (blue) galaxies. Bins, lines, and shaded regions are the same as in Fig.~\ref{fig:sAM_SMplane}. Surprisingly, for a given stellar mass, bulge-dominated galaxies have a higher $j_{\mathrm{dark matter}}$ than disk-dominated galaxies, while intermediate galaxies have the lowest $j_{\mathrm{dark matter}}$.}\label{fig:DMsAM_SMplane}
\end{figure}

\citet{Cole1996}, \citet{Macci2007}, and \citet{Knebe2008} showed that there is little-to-no correlation between the \citep{Bullock2000} halo spin parameter and halo mass at $z=0$. Nonetheless, there is an existing correlation between $j_{\mathrm{dark matter}}$ and halo mass \citep{Fall1980, Fall1983}. We now examine the relation between $j_{\mathrm{dark matter}}$ and both stellar mass and halo mass for bulge-dominated, intermediate, and disk-dominated galaxies. Figure~\ref{fig:DMsAM_SMplane} shows the relationship between $j_{\mathrm{dark matter}}$ and stellar mass for our different morphological samples. We find that this relationship is complex. Surprisingly, at fixed stellar mass, bulge-dominated galaxies have higher $j_{\mathrm{dark matter}}$ than disk-dominated galaxies. Moreover, intermediate galaxies have the lowest $j_{\mathrm{dark matter}}$ out of the three galaxy populations. Disk-dominated and intermediate galaxies have a $j_{\mathrm{dark matter}}$ scatter of 0.4 and 0.7 dex, respectively, both increasing with stellar mass to one dex. Bulge-dominated galaxies have a scatter of 0.7 dex, slightly decreasing with stellar mass. We tested our results against the earlier SAGE model \citep{Croton2016}. Although the disk-to-total stellar mass morphology distribution is different, the morphology sequence in $j_{\mathrm{dark matter}}$ is consistent.

\begin{figure}[t]
\centering
\includegraphics[width=\columnwidth, clip]{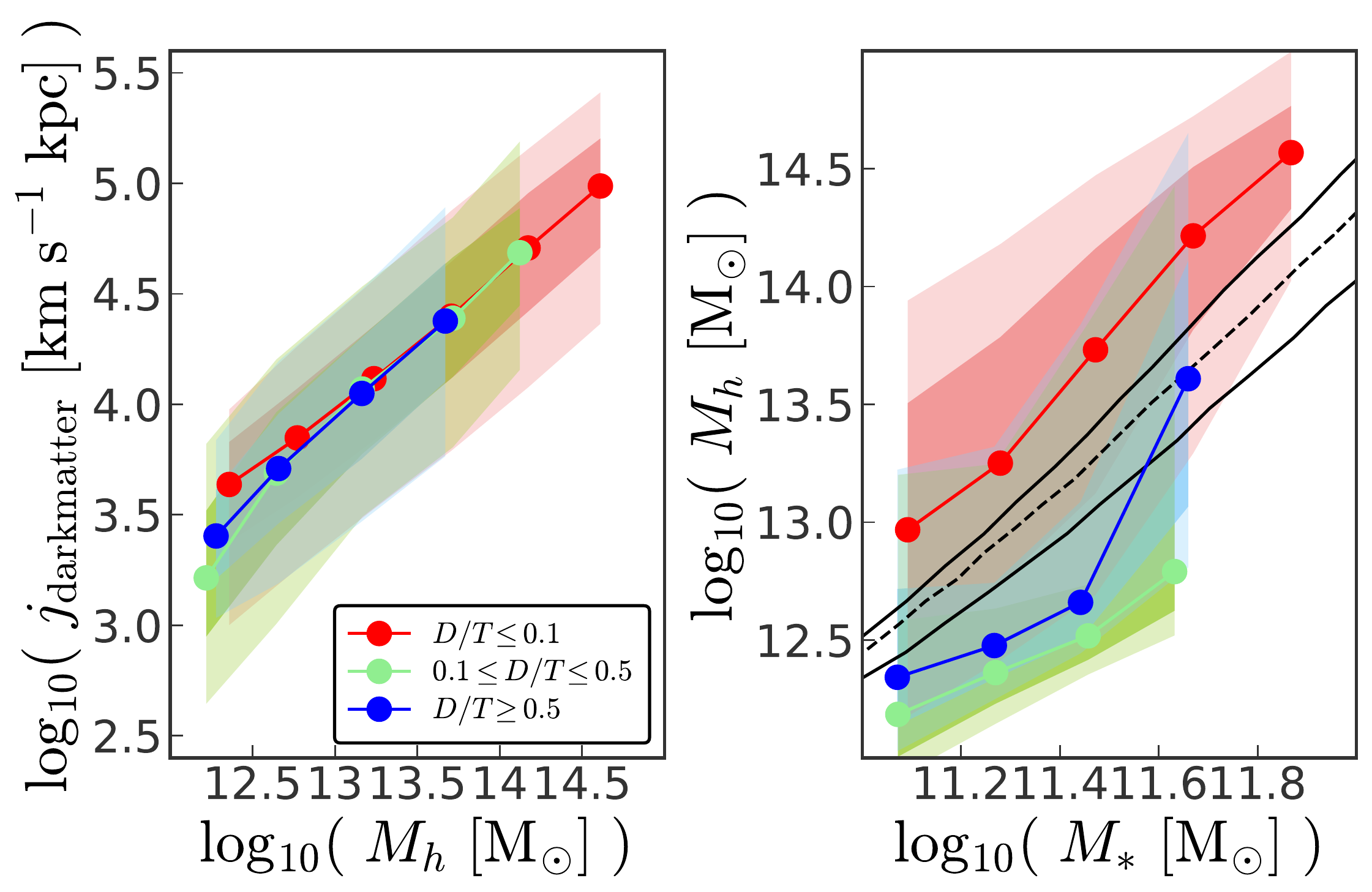}
\caption{Left panel: Dark matter specific angular momentum as a function of halo mass. Right panel: Dark matter halo mass as a function of stellar mass. Bins, colored lines, and shaded regions are the same as in Fig.~\ref{fig:sAM_SMplane}. The black dashed and solid lines show the best-fitting relation and one-sigma scatter, respectively, from the abundance matching result of \citet{Kravtsov2014StellarHalos}. Bulge-dominated galaxies live in halos that are about five to ten times more massive than disk-dominated galaxies, at fixed stellar mass, which explains the result in Fig.~\ref{fig:DMsAM_SMplane}. However, there is no dependence of morphology on $j_{\mathrm{dark matter}}$ when controlling for halo mass.}\label{fig:SMHMR_morph}

\end{figure}

To understand how disk-dominated galaxies can have a higher $j_{\mathrm{stellar}}$ (Fig.~\ref{fig:sAM_SMplane}), but lower $j_{\mathrm{dark matter}}$ than bulge-dominated galaxies (Fig.~\ref{fig:DMsAM_SMplane}), we explore the contribution of halo mass to the overall angular momentum. The left panel of Figure~\ref{fig:SMHMR_morph} shows $j_{\mathrm{dark matter}}$ as a function of halo mass for bulge-dominated (red), intermediate (green), and disk-dominated (blue) galaxies. When controlling for halo mass, the relationship between angular momentum and morphology vanishes entirely. This means that halo mass is the main contributor to the morphological trends seen in Figure~\ref{fig:DMsAM_SMplane},  while $j_{\mathrm{dark matter}}$ does not seem to be driving galaxy morphology.

The right panel of Figure \ref{fig:SMHMR_morph} shows the stellar mass-to-halo mass relation for all three galaxy populations in our sample. The black dashed line comes from \citet{Kravtsov2014StellarHalos}, who used the \citet{Bernardi2013TheProfile} stellar mass function to obtain abundance matching results. The one-sigma scatter is shown by the black solid lines. We find that, at fixed stellar mass, bulge-dominated galaxies have higher halo masses than disk-dominated galaxies, while intermediate galaxies live in the lowest mass halos. The halo mass scatter for disk-dominated and intermediate galaxies is about 0.5 dex increasing with stellar mass, while for bulge-dominated galaxies is about 1.0 dex, decreasing with stellar mass to 0.5 dex. 
A close examination of these results shows that bulge-dominated galaxies live in halos that are about five to ten times more massive than the ones that disk-dominated galaxies live in. As we explore in the next section, bulge-dominated galaxies formed from halo mergers and their high $j_{\rm dark matter}$ comes purely from their high mass, which is expected for higher merger rates \citep{Toomre1977, White1978, Heyl1994, Barnes1996}. 

\subsection{Contribution of bulge components to angular momentum}

\begin{figure}[t]
\centering
\includegraphics[width=1.1\columnwidth, clip]{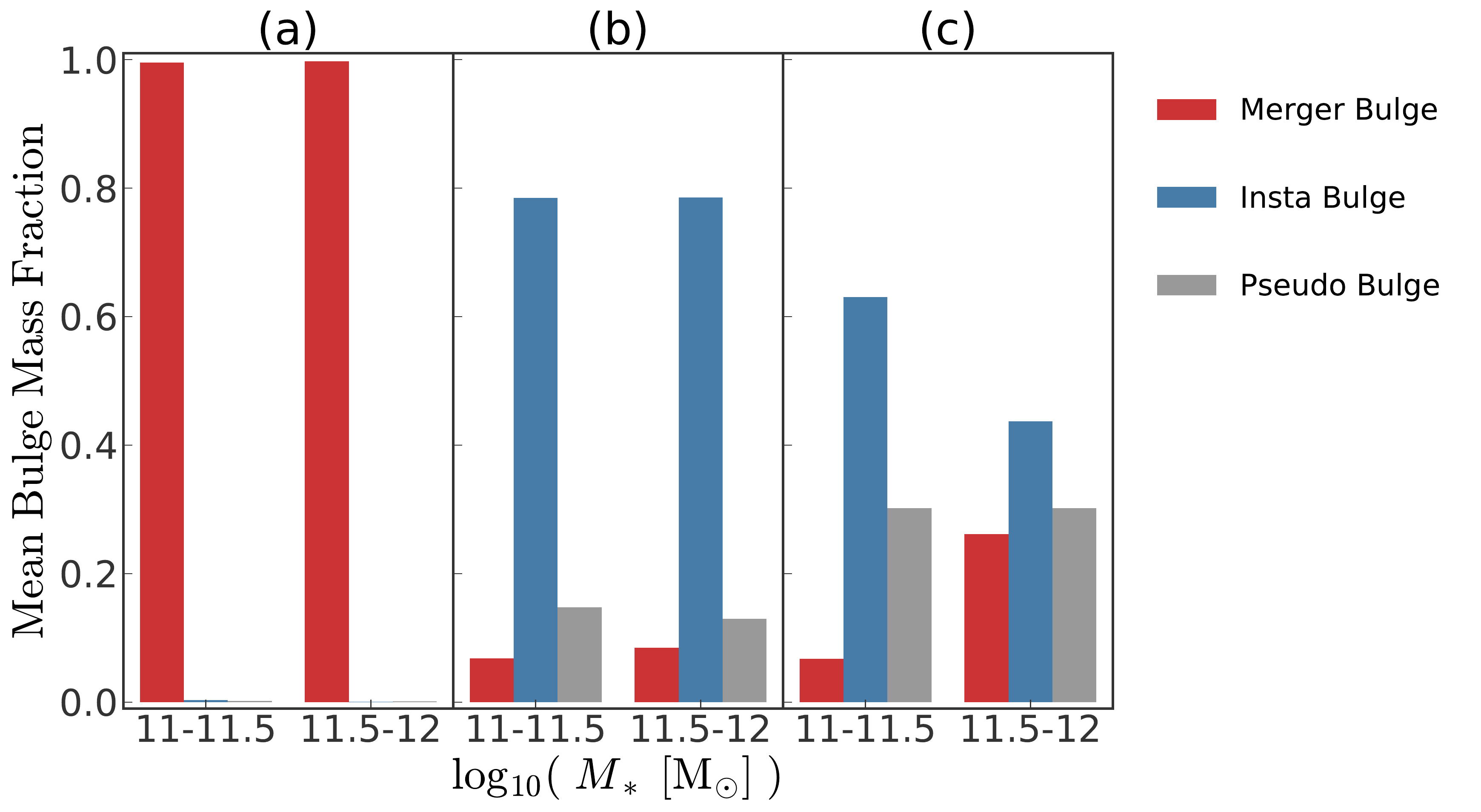}
\caption{Fraction of merger-driven (red), instability-driven (blue), and pseudo-bulge (grey) mass over the total bulge mass within the stellar ranges of $10^{11}-10^{11.5}$ and $10^{11.5}-10^{12} \mathrm{M}_{\odot}$ for bulge-dominated (region a), intermediate (region b), and disk-dominated galaxies (region c). Most of the mass in the bulge for bulge-dominated galaxies comes from mergers. For intermediate and disk-dominated galaxies, the instability-driven bulge contributes the most to their bulge mass.}\label{fig:bulgefrac_SMplane}
\end{figure}

The scatter in {\sc Dark Sage} galaxies within our sample encodes the diversity of galaxy formation pathways, some which are linked to different bulge formation channels. As outlined in Section \ref{galaxymorphology}, {\sc Dark Sage} galaxies have three components that contribute to their total bulge stellar mass. To distinguish the different kinds of bulges that contribute to multiple mechanisms to form galaxies, Figure~\ref{fig:bulgefrac_SMplane} shows the fractions of merger-driven (red), instability-driven (blue), and pseudo-bulge (grey) mass over the total bulge mass within the stellar mass ranges of $10^{11}-10^{11.5}$ and $10^{11.5}-10^{12} \mathrm{M}_{\odot}$ for bulge-dominated (Region a), intermediate (Region b), and disk-dominated galaxies (Region c). We find that mergers contribute 99.9\% of the mass of the bulge in bulge-dominated galaxies (Region a). For both intermediate and disk-dominated galaxies, the instabilities drive about 80 and 60 percent of the total bulge mass in their system, respectively.

To explore what contributes to the bulge build-up, we also looked at the time of last major merger for all galaxies. We find that 80\% of bulge-dominated galaxies had at least one major merger in the last 5.5~Gyr. In contrast, about 80\% of intermediate and disk-dominated galaxies did not experience a major merger, and of those that had major mergers, 20\% percent had their most recent merger more than 5.5~Gyr ago. After a major merger, the angular momentum of every component can change significantly \citep{Bekki1998, Cox2006, Lotz2010TheMergers, Moreno2015MappingFormation, Sparre2016ZoomingSimulations}. Much of this change depends on the gas fraction of the merging galaxies. If the smaller merging galaxy has no gas, then any gas disk in the larger system will persist relatively unimpeded \citep{Del2018}. For a merger of gas rich galaxies, most of the gas is consumed in the starburst and black hole accretion triggered by the merger. Minor mergers will also disrupt the structure of the gas disk, causing star formation in various parts of the disk, which could alter $j_{\mathrm{stellar, disk}}$. In essence, a large number of mergers can translate to a diversity of angular momentum.

\section{Summary and Discussion}\label{discussion}

We used the {\sc Dark Sage} semi-analytic model to investigate the relationship between the stellar mass, morphology, and specific angular momentum of galaxies with their parent halo properties. {\sc Dark Sage} is unique in its treatment of disk evolution, as it breaks each disk into a series of annuli of fixed specific angular momentum. Crucially, the net baryonic angular momentum of a galaxy depends on the entire history of its halo, and can therefore decouple from the halo's instantaneous properties.  This not only has downstream effects for how the disks and bulges of galaxies are grown, but it also means the model is predictive (as opposed to prescribed) when it comes to relating galaxy morphology to the halo at z=0.  We have explored exactly this in this paper for galaxies with stellar masses between $ 10^{11}-10^{12} \mathrm{M}_{\odot}$. Our results are summarized as follows:

\begin{itemize}
    \item {\sc Dark Sage} can reproduce the observational $j_{\mathrm{stellar, disk}}$, morphology, and stellar mass relation. We show a morphological sequence where at fixed stellar mass, $j_{\mathrm{stellar, disk}}$ increases with $D/T$ (Fig.~\ref{fig:sAM_SMplane}). This relation qualitatively follows several observational studies \citep{Fall1983, Romanowsky2012, Fall2013, Posti2018, Sweet2018, Fall2018ANGULARBULGES}.
    \item {\sc Dark Sage} predicts that at fixed stellar mass, galaxies with high $j_{\mathrm{total, disk}}$ (stars and cold gas) disk tend to be bulge-dominated (Fig.~\ref{fig:BaryonicAM_SMplane}). Bulge-dominated galaxies have disks with higher gas fractions. Generally, gas disks tend to have higher \textit{j} than stellar disks, as disk stars preferentially form from the lowest-\textit{j} gas in the disk. Thus, the higher a disk's gas fraction, the higher its \textit{j}. We also show that the $j_{\mathrm{total, disk}}$ of disk-dominated and intermediate galaxies traces the $j_{\mathrm{stellar, disk}}$.
    \item We find that the relationship between $j_{\mathrm{dark matter}}$ and stellar morphology is not simple. At fixed stellar mass, bulge-dominated galaxies have a higher $j_{\mathrm{dark matter}}$ than disk-dominated galaxies. Intermediate galaxies have the lowest $j_{\mathrm{dark matter}}$ (Fig.~\ref{fig:DMsAM_SMplane}). 
    \item To understand how bulge-dominated galaxies have high $j_{\mathrm{dark matter}}$, we explore the correlation between the $j_{\mathrm{dark matter}}$ and halo mass. We find that halo mass is the main contributor to the morphological trends with $j_{\mathrm{dark matter}}$. When controlling for halo mass, the relationship between angular momentum and morphology vanishes (left panel of Fig.~\ref{fig:SMHMR_morph}). We also find that Bulge-dominated galaxies live in halos that are about five to ten times more massive than disk-dominated galaxies, whereas intermediate galaxies live in the least massive halos (right panel of Fig.~\ref{fig:SMHMR_morph}). Based on these results, halo mass and not $j_{\mathrm{dark matter}}$ is the driver of the  morphological trend we see.
\end{itemize}

It is important to point out that the results in this paper focus on morphological, rather than spectro-photometric, trends. When using the specific star formation rate instead of $D/T$, our results for the $j_{\mathrm{stellar, disk}}$ sequence were in agreement. We find the same color sequence, where active galaxies have the highest $j_{\mathrm{stellar, disk}}$ followed by intermediate and passive galaxies. Additionally, we find that passive galaxies have a higher $j_{\mathrm{dark matter}}$ than active and intermediate galaxies. However, in this case, there is a color-sequence for $j_{\mathrm{dark matter}}$ that differs from the morphological sequence. At fixed stellar mass, intermediate galaxies have higher $j_{\mathrm{dark matter}}$ than active galaxies, but lower $j_{\mathrm{dark matter}}$ than passive galaxies. Here, active galaxies have the lowest $j_{\mathrm{dark matter}}$. These results lead us to believe that intermediate galaxies from the stellar morphology definition are not green valley galaxies. Results within our sample show a fundamental distinction between morphology and specific star formation rate for intermediate galaxies.

Our results imply that the angular momentum of the disk and that of dark matter are not tied in massive galaxies, given that {\sc Dark Sage} consistently treats angular momentum evolution. In {\sc Dark Sage}, the size of the disk and the halo spin are correlated, but not bound to each other (this is reflected, for example, in the result of HI-excess galaxy by \citealt{Lutz2018}). This relationship has serious implications for our understanding of galaxy morphology and color. We have shown that the angular momentum of the disk and dark matter differs for each galaxy type. The halo spin in {\sc Dark Sage} determines how gas in a single cooling episode is added to the disk. The way in which matter is distributed in the disk depends on the galaxy's entire history. This history is tied to the halo spin evolution, which relates to the amount of gas being ejected from feedback, star formation and merger history as well as other physical mechanisms. 

Our results also show that massive galaxies have a scatter larger than 0.2 dex at fixed halo mass \citep{Behroozi2010}. The scatter in the stellar mass-to-halo mass relation has profound implications on the galaxy-halo connection. Understanding the scatter may constrain different galaxy quenching models that describe diverse ways in which galaxies gain halo mass \citep{Behroozi2019UniverseMachine:10, Man2019THEPROPERTIES}. Future work will examine the relationship between halo mass, morphology, and color by looking at the stellar mass-to-halo mass relation to further understand the morphological sequence found in this paper.

\acknowledgments
\textbf{ACKNOWLEDGEMENTS}
We thank Amanda Moffett, Darren Croton, and Victor Calder\'on for insightful conversations. This paper used data from the Theoretical Astrophysical Observatory (TAO) \url{https://tao.asvo.org.au/}. TAO is part of the All-Sky Virtual Observatory (ASVO) funded and supported by Swinburne University of Technology, Astronomy Australia Limited, and the Australian Government through the Commonwealth's Education Investment Fund and National Collaborative Research Infrastructure Strategy. We used computational facilities from the Vanderbilt Advanced Computing Center for Research and Education (ACCRE). Literature reviews for this work was made using the NASA’s Astrophysics Data System. Results were produced using the IPython package \citep{Perez2007IPythonFor}, Scipy \citep{Jones2001SciPy:Python}, matplotlib \citep{Hunter2007ComputingEngineering}, Astropy \citep{Collaboration2013AstrophysicsAstronomy}, and NumPy \citep{VanDerWalt2011TheComputation}.

\clearpage

\bibliographystyle{aasjournal}
\bibliography{bibfile.bib}

\end{document}